\documentclass[aip,jcp,amsmath,amssymb,amsfonts,groupedaddress,floatfix,reprint]{revtex4-1}
\usepackage{graphicx}
\usepackage{dcolumn} 
\usepackage{ifthen}

\newcommand{\nn}[0]{\nonumber \\}
\ifthenelse{\isundefined{\vr}}{ \newcommand{\vr}[0]{ {\boldsymbol{r}} } }{ \renewcommand{\vr}[0]{ {\boldsymbol{r}} } }
\newcommand{\bra}[0]{\langle}
\newcommand{\ket}[0]{\rangle}
\newcommand{\ks}[0]{\mathrm{s}}
\newcommand{\xc}[0]{\mathrm{xc}}
\newcommand{\exch}[0]{\mathrm{x}}
\newcommand{\corr}[0]{\mathrm{c}}
\newcommand{\unif}[0]{\mathrm{unif}}

\newcommand{\tLDA}[0]{\mathrm{tLDA}}
\newcommand{\fictitious}[0]{\mathrm{f}}
\newcommand{\Hartree}[0]{\mathrm{H}}
\newcommand{\TF}[0]{\mathrm{TF}}
\newcommand{\occ}[0]{\mathrm{occ}}
\newcommand{\ind}[2]{ \int\!\!\mathrm{d}^{#1}#2 \; }

\begin{document}

\title{ Derivative discontinuity and exchange-correlation potential \\ of meta-GGAs in density-functional theory }

\author{F. G. Eich}
\email[]{eichf@missouri.edu}
\affiliation{Department of Physics, University of Missouri-Columbia, Columbia, Missouri 65211, USA}

\author{Maria Hellgren} 
\affiliation{International School for Advanced Studies (SISSA), via Bonomea 265, 34136 Trieste, Italy}

\begin{abstract}
  We investigate fundamental properties of meta-generalized-gradient approximations (meta-GGAs) to the 
  exchange-correlation energy functional, which have an implicit density dependence via the Kohn-Sham kinetic-energy density. 
  To this purpose, we construct the most simple meta-GGA by expressing the local exchange-correlation energy per
  particle as a function of a fictitious density, which is obtained by inverting the Thomas-Fermi kinetic-energy
  functional. This simple functional considerably improves the total energy of atoms as compared to the standard
  local density approximation. The corresponding exchange-correlation potentials are then determined exactly through
  a solution of the optimized effective potential equation. These potentials support an additional
  bound state and exhibit a derivative discontinuity at integer particle numbers. We further demonstrate that through
  the kinetic-energy density any meta-GGA incorporates a derivative discontinuity. However, we also find that for commonly
  used meta-GGAs the discontinuity is largely underestimated and in some cases even negative. 
\end{abstract}


\date{\today}

\maketitle

\section{Introduction} \label{SEC:Introduction}

Density-functional theory (DFT) is a well-established method for computing the electronic structure
of atoms, molecules and solids.\cite{HK,parrbook,grossbook} DFT provides an 
exact mapping between interacting electrons and noninteracting Kohn-Sham (KS) electrons by requiring
that the electronic density ${n(\vr)}$ is identical in both systems.\cite{KS} This so-called KS scheme
allows for an efficient computation of total energies. The accuracy of the approach depends, however, on the elusive exchange-correlation (xc) energy 
functional $E_\xc[n]$, which contains all the effects of the electron-electron interaction beyond 
the classical Hartree approximation. The functional derivative of $E_\xc[n]$ with respect to the density 
yields the xc potential $v_\xc[n](\vr)$, which is part of the local potential acting on the KS electrons.
Its dependency on the density is, however, nonlocal and incorporating this nonlocality is a challenge 
for density functional approximations.

Starting from the local density approximation (LDA), there has been an enormous effort in constructing successively 
better approximations to $E_\xc[n]$.\cite{functionaloverview}
These approximations are usually categorized by their respective ingredients: Semi-local approximations, which depend 
on the gradient of the density, are referred to as generalized gradient approximations (GGAs). 
An additional dependence on the KS kinetic-energy density leads to so-called meta-GGAs, and the 
inclusion of exact exchange leads to hybrids or hyper-GGAs. The kinetic-energy density and
the exchange energy are given in terms of occupied KS orbitals. At the highest level of sophistication many-body perturbation theory~\cite{fetterbook}
(MBPT) is used to derive approximations based on the KS Green's function. These approximations depend not only
on occupied KS orbitals, but also on unoccupied KS orbitals and the KS single-particle energies.\cite{vonbarth}
Meta-GGAs, hyper-GGAs and MBPT-based functionals are implicit density functionals, since they 
depend on the density indirectly via KS orbitals and energies. Therefore these functionals develop a
nonlocal dependence on the density.

For implicit density functionals the so-called optimized effective potential (OEP) equation has to be solved
in addition to the KS equation.\cite{talshad,gkkg,ak03,kkreviewOEP} The OEP equation is numerically challenging
but has shown to produce very accurate xc potentials.\cite{engel,hvb07} It can be shown that the xc potential
of MBPT-based functionals generate a density approximating the density coming from the nonlocal self-energy
associated with the corresponding MBPT functional.\cite{casida95}
And, as will be shown in this work, the OEP equation for meta-GGAs yields a local xc potential that produces a 
density approximating the density coming from a Hamiltonian with a position-dependent mass in
the kinetic-energy operator. Although the density of the OEP KS system is similar to the density of
the corresponding nonlocal or position-dependent mass Hamiltonian, the orbitals and orbital energies are,
in general, different.

The energy of the highest occupied molecular orbital (HOMO) in the KS system corresponds to
the true ionization energy of the interacting system.\cite{Almbladh} The true fundamental gap, however, does 
not coincide with the fundamental gap of the KS system. The KS gap, given by the energy 
difference of the lowest unoccupied molecular orbital (LUMO) and the HOMO of the KS system,
has to be corrected by the so-called derivative discontinuity, which appears as a kink in the xc energy and as a constant shift in the xc potential 
when crossing integer particle numbers.\cite{pplb82} Being explicit density functionals, the LDA and 
GGAs do not show a derivative discontinuity,\footnote{Note that in Ref.\ \onlinecite{ak13} a GGA has been
proposed which exhibits a derivative discontinuity.}
which means that the fundamental gap is usually underestimated. In contrast to this, implicit density
functionals exhibit a derivative discontinuity which is given by the difference of the HOMO-LUMO gap in
the KS system and the HOMO-LUMO gap in the corresponding MBPT Hamiltonian or--in the case of meta-GGAs--the Hamiltonian with a 
position-dependent mass. The former has been studied extensively in the literature,\cite{kli,hg12,hg13} 
while the latter has, so far, not received much attention. 

An inclusion of the derivative discontinuity is becoming increasingly more important for functional constructions: 
For example, it has been shown that the derivative discontinuity is crucial for the breaking of certain chemical bonds,\cite{pplb82,perde,krokum} to 
capture charge transfer excitations\cite{tozer,hg12,hg13} and for describing strongly correlated materials.\cite{yang} 
We will here demonstrate that the derivative discontinuity is present 
in any meta-GGA and derive a compact equation to determine its size. 

Several different advanced meta-GGAs have been constructed in recent years.\cite{vvs98,tpss,m06l,tpssrev} 
The additional kinetic-energy density dependence has shown to improve ground-state properties considerably.
Furthermore, it has shown some promise in calculating the optical spectra of semiconductors.\cite{nv11} 
In this work we propose a simple model for a meta-GGA, which is based on an idea first 
proposed by Ernzerhof and Scuseria.\cite{es99} This model meta-GGA, dubbed tLDA, is constructed 
by replacing the density dependence in the LDA with a kinetic-energy density dependence
via the Thomas-Fermi relation of the electron gas. The tLDA improves the total energy of atoms in comparison to the standard LDA.
We will use the tLDA to investigate fundamental properties of meta-GGA functionals and compare its local xc potential and derivative discontinuity
to two popular meta-GGAs, the Tao-Perdew-Staroverov-Scuseria (TPSS)~\cite{tpss} and van Voorhis and Scuseria (VS98)~\cite{vvs98} approximation.

The paper is organized as follows: We introduce the tLDA in Sec.\ \ref{SEC:tLDA}. In Sec.\ \ref{SEC:mGGA_OEP}, we derive the OEP equation for a 
generic meta-GGA potential and propose a physical interpretation of the kinetic-energy density dependence in terms of a position-dependent mass. 
In Sec.\ \ref{SEC:mGGA_OEP_KLI}, we employ the KLI approximation to the OEP equation which allows us to anticipate the derivative discontinuity.
Our main formal result, i.e., the derivation of the derivative discontinuity for meta-GGAs, is obtained in Sec.\ \ref{SEC:DD} where we also provide a simple and physically
transparent equation determining its magnitude. In Sec.\ \ref{SEC:results}, we present numerical solutions to the full OEP equation for a set of
spherical atoms using the tLDA as well as the TPSS and VS98 approximations. Our conclusions are presented in Sec.\ \ref{SEC:conclusions}.

\section{A model meta-GGA} \label{SEC:tLDA}

\begin{table*}[t]
  \caption{Total energies (Hartree atomic units) of spherical atoms in different approximations at selfconsistent densities.}
  \begin{ruledtabular}
    \begin{tabular}{crrrrrrrrr}
      Atom & LDAx\footnotemark[1] & LDA & tLDAx\footnotemark[1] & tLDA & PBE & TPSS & VS98 & EXX & Exp.\footnotemark[2] \\
      \hline
      He & -2.723 & -2.835 & -2.765 & -2.882 & -2.893 & -2.910 & -2.917 &-2.862 &-2.904 \\
      Be & -14.223 & -14.447 & -14.334 & -14.569 & -14.630 & -14.672 & -14.696 & -14.572&-14.667 \\
      Ne & -127.490 & -128.233 & -127.922 & -128.685 & -128.866 & -128.981 & -129.020 &-128.545& -128.938 \\
      Mg & -198.248 & -199.139 & -198.805 & -199.723 & -199.955 & -200.093 & -200.163 & -199.612&-200.053 \\
      Ar & -524.517 & -525.946 & -525.461 & -526.922 & -527.346 & -527.569 & -527.747 &-526.812 &-527.540
    \end{tabular}
  \end{ruledtabular}
  \footnotetext[1]{ Evaluated at selfconsistent xc density. } 
  \footnotetext[2]{ From Refs.\ \onlinecite{atomicEnergies1, atomicEnergies2}. }
  \label{TAB:totalEnergyXC}
\end{table*}
In its usual implementation DFT maps interacting electrons onto noninteracting KS electrons. 
Within KS DFT the total energy can be written as
\begin{align}
  E_{v}[n] = T_\ks[n] + \ind{3}{r} v(\vr) n(\vr) + E_\Hartree[n] + E_\xc[n] ~, \label{Etot}
\end{align}
where $v(\vr)$ is the external potential and $T_\ks[n]=\ind{3}{r} \tau(\vr) $ the kinetic energy.
The KS kinetic-energy density, here defined in its symmetric form, is given by 
\begin{align}
  \tau(\vr) = \frac{1}{2} \sum_i^\occ \left|\nabla \phi_i(\vr) \right|^2 ~,\label{tauDefinition}
\end{align}
with $\phi_i$ being KS orbitals (Hartree atomic units are used throughout the paper). The Hartree energy is given by
\begin{align}
  E_\Hartree[n] = \frac{1}{2} \ind{3}{r}\!\mathrm{d}^3r' \frac{n(\vr) n(\vr')}{|\vr-\vr'|} ~, \label{EHartree}
\end{align}
and $E_\xc[n]$ is the xc energy functional, which has to be approximated. 

A precursor of DFT is Thomas-Fermi (TF) theory. In the context of DFT one can view TF theory as minimizing the total energy functional, Eq.\ \eqref{Etot}, directly. 
This, however, requires an explicit approximation for the kinetic energy in terms of the density. The TF kinetic-energy functional
is the LDA for $T_\ks[n]$. It uses the kinetic-energy density for the noninteracting uniform electron gas, which can be written 
explicitly in terms of the density, i.e.,
\begin{align}
  \tau^\unif(n) = C_\TF n^{5/3} = \tfrac{3}{10} (3 \pi^2 n)^{2/3} n ~. \label{TFrelation}
\end{align}
Furthermore, the exchange energy per particle of the uniform electron gas is given by 
\begin{align}
  \epsilon_\exch^\unif(n) = -C_\exch n^{1/3} = -\tfrac{3}{4 \pi}\left(3 \pi^2 n \right)^{1/3} ~.  \label{exxLDA}
\end{align}
Also the correlation energy per particle $\epsilon_\corr^\unif(n)$ is available thanks to accurate parametrizations~\cite{vwn,pw} of 
quantum Monte-Carlo calculations for the interacting electron gas.\cite{ceperley,ballone} 
Accordingly, the total energy within TF theory can be written as
\begin{align}
  E_v^\TF[n] & = \ind{3}{r} n(\vr) \Big\{ C_\TF n^{2/3}(\vr) + v(\vr) \nn
  & \phantom{ = \ind{3}{r} n(\vr) \Big\{} {} + \tfrac{1}{2} v_\Hartree(\vr)  + \epsilon_\xc^\unif[n(\vr)] \Big\} ~, \label{TFenergy}
\end{align}
where we combine the LDA for exchange and correlation to the xc energy per particle, $\epsilon_\xc^\unif(n) = \epsilon_\exch^\unif(n) + \epsilon_\corr^\unif(n)$,
and $v_\Hartree(\vr)= \ind{3}{r'}n(\vr')/|\vr-\vr'|$ is the Hartree potential. It is a well-known fact that for an inhomogeneous system the TF approximation is missing 
fundamental quantum features such as the atomic shell structure. The usual implementation of DFT builds these features into the theory by employing the KS scheme, 
which treats $T_\ks[n]$ exactly.

\begin{figure}[b]
  \includegraphics[width=8.5cm, clip=true]{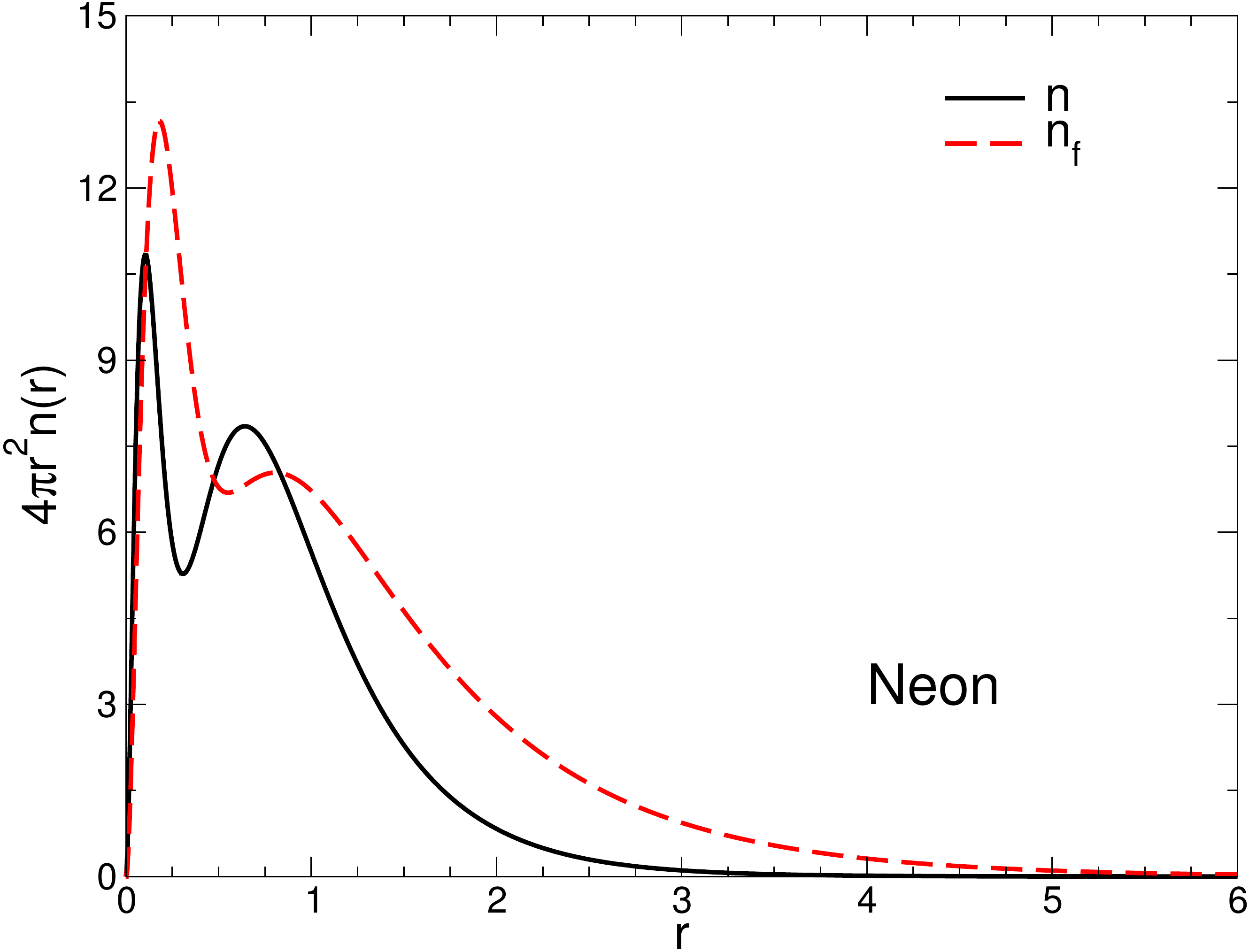}
  \caption{(Color online) The density and the fictitious density, defined in Eq.\ \eqref{nf}, for Neon evaluated at the LDA density.}
  \label{FIG:fictitiousDensity}
\end{figure}

The kinetic-energy density is given in terms of the KS orbitals which are implicit functionals of the density via the KS potential.
The dependency on the density is therefore generally nonlocal, because a variation of the density at a given point $r$, which, in turn affects
the KS potential, will produce a change in the orbital at point $r'$. 

In order to build this nonlocality into the xc energy functional, we use
an idea first proposed by Ernzerhof and Scuseria.\cite{es99}
They used the TF relation \eqref{TFrelation} ``backwards'' to replace the density dependence in the LDA xc functional with a kinetic-energy density dependence. 
While in Ref.\ \onlinecite{es99} the full density dependence has been replaced, we only employ this replacement in the energy density per particle,
thus generating a functional depending on, both, $n$ and $\tau$. The reason for using the kinetic-energy density only in the energy per particle is
that this functional (tLDA) can still be interpreted as an approximation for the averaged xc hole using the xc hole of the uniform electron gas. 
Accordingly, the averaged xc hole of the tLDA obeys important sum rules, since it is taken from a physical reference system.
The only difference to the LDA is that the density of the uniform electron gas, from which the xc hole is taken,
is determined by $\tau$ via Eq.\ \eqref{TFrelation} and not $n$. Explicitly, the tLDA is given by
\begin{align} 
  E^\tLDA_\xc[n] & = \ind{3}{r} n(\vr) \tilde{\epsilon}_\xc^\unif[\tau(\vr)] \nn
  & = \ind{3}{r} n(\vr) \epsilon_{\xc}^{\unif}[n_\fictitious(\vr)] ~, \label{tLDA}
\end{align}
where,
\begin{align}
  n_\fictitious(\vr) = \left( C_\TF^{-1} \tau(\vr) \right)^{3/5} =\frac{1}{3 \pi^2} \left(10 \pi^2 \tau(\vr)\right)^{3/5} ~, \label{nf}
\end{align}
is the ``fictitious'' density determining the density of the interacting electron gas from which the xc hole is borrowed.
The tLDA is thus based on the same reference system as the usual LDA, which means that both 
are exact in the limit of a constant density. However, only the LDA is an explicit and local density functional. 
The tLDA belongs to the class of meta-GGA functionals due to the dependence on $\tau$. The relation
between the LDA and the tLDA can be summarized as follows: In the LDA the xc hole is taken to be consistent with
the density, while in the tLDA the xc hole is taken to be consistent with the kinetic-energy density.

In Fig.\ \ref{FIG:fictitiousDensity} we plot the density and the fictitious density of the Neon atom.
We notice that the fictitious density has smoother shell oscillations than the true density. It also has a slower asymptotic decay
and does not necessarily integrate to the number of particles. In Table \ref{TAB:totalEnergyXC} we present total energies
for a number of spherically symmetric atoms. Despite the simplicity of our construction the total energy in tLDA is
significantly improved as compared to LDA and the main improvement can be found in the exchange part (tLDAx). 
Compared to exact exchange (EXX) the tLDAx exchange energy is still underestimated but in both LDA and tLDA the overestimated 
correlation energy compensates this error leading to a reasonable total energy. The improved exchange energy 
within tLDA, compared to LDA, reduces the error in the total energy by almost half.

\section{Kohn-Sham potential of meta-GGAs } \label{SEC:mGGA_OEP}

An xc functional with a dependence on the kinetic-energy density has an implicit dependency on the density 
via the KS potential that generates the KS orbitals. Accordingly, the OEP scheme has to be employed in 
order to obtain the corresponding xc potential. In this section we will derive the OEP equation for 
a generic meta-GGA and provide a physical interpretation.

\subsection{General derivation} \label{SEC:mGGA_OEP_generic}

The KS potential $v_\ks$ is usually decomposed into 
\begin{align}
  v_\ks(\vr) = v(\vr) + v_\Hartree(\vr) + v_\xc(\vr) ~, \label{vs}
\end{align}
where $v$ is the external potential and $v_\Hartree$ the Hartree potential.
The xc potential, $v_\xc$, is obtained from the functional derivative of the
xc energy functional with respect to the density, i.e.,
\begin{align}
  v_\xc(\vr) = \frac{\delta E_\xc[n]}{\delta n(\vr)} ~. \label{vxc}
\end{align}

In the following we will consider a generic meta-GGA, written in the form
\begin{align}
  E_\xc[n] = \ind{3}{r} n(\vr) \epsilon_\xc[n(\vr),\nabla n(\vr),\tau(\vr)] ~. \label{excMGGA}
\end{align}
Thus, $E_\xc[n]$ has a dependency on the density, its gradient, and the kinetic-energy density as defined in Eq.\ \eqref{tauDefinition}.
When evaluating the functional derivative of Eq.\ \eqref{vxc} it is convenient to split the xc potential into two terms:
$v_{\xc,1}$ coming from the explicit dependence on the density and its gradient,
\begin{align}
  v_{\xc,1}(\vr)= \epsilon_\xc(\vr) + n(\vr)\frac{\partial\epsilon_\xc}{\partial n}(\vr) - \nabla \cdot n(\vr) \frac{\partial  \epsilon_\xc}{\partial \nabla n}(\vr) ~, \label{vxc_1}
\end{align}
and ${v_{\xc, 2}}$ coming from the implicit density dependence via $\tau$, 
\begin{align}
  v_{\xc,2}(\vr) & = \ind{3}{r'} n(\vr') \frac{\partial \epsilon_{\xc}}{\partial \tau}(\vr') \frac{\delta \tau(\vr')}{\delta n(\vr)} \nn 
  & = \ind{3}{r'}\!\!\!\ind{3}{r''} \psi_\xc(\vr') \frac{\delta \tau(\vr')}{\delta v_s(\vr'')} \frac{\delta v_s(\vr'')}{\delta n(\vr)} ~. \label{vxc_2}
\end{align}
In the second line of Eq.\ \eqref{vxc_2} we have introduced the quantity
\begin{align}
  \psi_\xc(\vr) = \frac{\delta E_\xc}{\delta \tau(\vr)} = n(\vr)\frac{\partial \epsilon_{\xc}}{\partial \tau}(\vr) ~, \label{psi_xc}
\end{align}
and we have used the chain rule to evaluate the change of the kinetic-energy density due to a variation in the density.
$\delta \tau/\delta v_\ks$ is the kinetic-energy density--density response function, ${\chi_{\ks, \tau n}}$, which describes the change of
the kinetic-energy density due to a variation of the local potential. In terms of KS orbitals it is given by
\begin{align}
  \chi_{\ks, \tau n}(\vr, \vr') & = \frac{1}{2} \sum_{i \neq j} \frac{f_i -f_j}{\varepsilon_i - \varepsilon_j} \nn
  & \times {} \left[ \nabla \phi_i^{\star}(\vr) \right] \cdot \left[\nabla \phi_j (\vr) \right] \phi_j^{\star}(\vr') \phi_i(\vr') ~. \label{chi_tau_n}
\end{align}
Similarly, $\delta v_\ks/\delta n$ is the inverse of the density--density response function ${\chi_{\ks, n n}}$ of the KS system. 
We can now rewrite Eq.\ \eqref{vxc_2} in the following compelling form,
\begin{align}
  \ind{3}{r'} \chi_{\ks, n n}(\vr, \vr') v_{\xc,2}(\vr') = \ind{3}{r'} \chi_{\ks, n \tau}(\vr, \vr') \psi_\xc(\vr') ~, \label{OEP_mGGA}
\end{align}
where we used the reciprocity relation ${\chi_{n \tau}(\vr, \vr') = \chi_{\tau n}(\vr', \vr)}$. 
Equation \eqref{OEP_mGGA} allows us to interpret the potential ${v_{\xc,2}}$ as the local potential 
that yields the same density--to first order--as a local change in the electronic mass. In order to see this more clearly 
we consider a spatially varying mass ${\delta m}$ and add this to the KS equation. The kinetic-energy operator is then change by 
\begin{align}
  \hat{T}_\ks + \delta \hat{T}_\ks & = - \frac{1}{2} \nabla \cdot \frac{1}{1 + \delta m(\vr)} \nabla \nn
  & = - \frac{1}{2} \nabla^2 - \frac{1}{2} \nabla \cdot \psi_\xc(\vr) \nabla , \label{T_varying_mass}
\end{align}
with ${\psi_\xc \approx -\delta m}$ and ${\delta \hat{T}_\ks = - \tfrac{1}{2} \nabla \cdot \psi_\xc\nabla}$.
The first-order density response obtained by applying ${\delta \hat{T}_\ks}$ is ${\delta n=\chi_{\ks, n \tau}\psi_\xc}$.
It follows that Eq.\ \eqref{OEP_mGGA} is the condition that the first-order density response, due to the perturbation
${-\tfrac{1}{2} \nabla \cdot \psi_\xc \nabla - v_{\xc,2}}$, is zero, i.e., there is no difference in the density, to first order, in going from the
KS Hamiltonian $H_{\mathrm{KS}} = - \tfrac{1}{2}\nabla^2 + v + v_{\xc,1} + v_{\xc,2}$ to the generalized KS (GKS) Hamiltonian 
$H_{\mathrm{GKS}} = -\tfrac{1}{2}\nabla^2 -\tfrac{1}{2} \nabla \cdot \psi_\xc \nabla + v + v_{\xc,1}$. This interpretation is familiar 
from the OEP EXX approach in which the nonlocal Hartree-Fock self-energy plays the same role as ${\delta \hat{T}_\ks}$.\cite{casida95}
We mention that a KS Hamiltonian with a spatially varying mass has recently been shown to appear naturally, when time-dependent DFT 
is extended to address thermoelectric phenomena.\cite{edvv}

\subsection{KLI approximation} \label{SEC:mGGA_OEP_KLI}

There are various ways to approximate the full OEP equation.\cite{slater,ceda,kli,staroverov} 
A common approximation within EXX is the so-called KLI approximation.\cite{kli} 
It allows for a partial analytic inversion of the KS response function. The crucial step is to set all energy differences in the 
denominator of $\chi_\ks$ to the same value $\mathcal{E}$, which is known as the common-energy-denominator approximation (CEDA).\cite{ceda}
Within the EXX approximation the same energy denominators appear on the right and left hand side of the OEP equation
and, hence, the actual value of the energy denominator is not important. The resulting expression has thus no dependence
on orbital energies and can be written in terms of occupied orbitals only. 

The right hand side of the OEP equation derived for the $\tau$-dependent functional has, again, the same energy 
denominators as the left hand side. It is therefore possible to do a CEDA approximation for meta-GGAs. We find
\begin{align}
  \chi_{\ks, n \tau}(\vr,\vr') & \approx \frac{1}{2\mathcal{E}} \sum_i f_i \phi_i(\vr) \left[ \nabla_{\vr'} \phi_i^\star(\vr') \right] \cdot \nn
  & \times \sum_j (1-f_j) \phi_j^\star(\vr) \nabla_{\vr'} \phi_j(\vr') + \mathrm{c.c.} ~. \label{mGGA_CEDA_1}
\end{align}
Using the completeness of the KS orbitals we can write Eq.\ \eqref{mGGA_CEDA_1} as
\begin{align}
  & \chi_{\ks,n \tau}(\vr,\vr') \approx \frac{1}{2 \mathcal{E}} \sum_i f_i \phi_i(\vr) \left[ \nabla_{\vr'} \phi_i^\star(\vr') \right] \cdot \nn
  & \times \nabla_{\vr'} \left[ \delta(\vr -\vr') - \sum_j f_j \phi_j^\star(\vr) \phi_j(\vr')\right] + \mathrm{c.c.} ~. \label{mGGA_CEDA_2}
\end{align}
Since $\chi_{\ks, n \tau}$ is integrated with $\psi_\xc$ we can rewrite the first term 
using a partial integration, which yields the correction of the kinetic-energy operator due to a 
position-dependent mass [cf.\ Eq.\ \eqref{T_varying_mass}]. In the KLI approximation only the diagonal elements ${i = j}$ in the
second term are retained, i.e., 
\begin{align}
  \ind{3}{r'} & \chi_{\ks,n \tau}(\vr,\vr') \psi_\xc(\vr') \approx \nn
  & \frac{2}{\mathcal{E}} \Bigg( \sum_i^\occ \left( \phi_i(\vr)\left[ - \frac{1}{4} \nabla \cdot \psi_\xc(\vr) \nabla \phi_i^\star(\vr) \right] + \mathrm{c.c.} \right) \nn
  & \phantom{\frac{1}{\mathcal{E}} \Bigg(} {} - \sum_i^\occ |\phi_i(\vr)|^2 \frac{1}{2}\left\bra \nabla \phi_i \right| \psi_\xc \left| \nabla \phi_i \right\ket \Bigg) ~. \label{mGGA_KLI}
\end{align}
Performing the same steps to approximate $\chi_{\ks, n n}$ we find the KLI approximation to the OEP equation for meta-GGAs:\cite{ak03}
\begin{align}
  & v_{\xc,2}(\vr) =  \label{kli} \\
  & \frac{1}{n(\vr)} \Bigg( \sum_i^\occ \left( \phi_i(\vr)\left[ - \frac{1}{4} \nabla \cdot \psi_\xc(\vr) \nabla \phi_i^\star(\vr) \right] + \mathrm{c.c.} \right) \nn
  & {} + \sum_i^\occ |\phi_i(\vr)|^2 \left[ \left\bra \phi_i \right| v_{\xc, 2} \left| \phi_i \right\ket - 
    \frac{1}{2}\left\bra \nabla \phi_i \right| \psi_\xc \left| \nabla \phi_i \right\ket \right] \Bigg) ~. \nonumber
\end{align}

The exact OEP equation \eqref{OEP_mGGA} as well as the KLI equation \eqref{kli} only determine $v_{\xc,2}$ 
up to an overall constant. In order to fix this constant we can exclude the HOMO in the summation over
occupied orbitals in the second term in the KLI equation \eqref{kli}. This is equivalent of setting 
\begin{align}
  0  = \frac{1}{2}\ind{3}{r} \psi_\xc(\vr) |\nabla\phi_\mathrm{H}(\vr)|^2 - \ind{3}{r} v_{\xc,2}(\vr) |\phi_\mathrm{H}(\vr)|^2 ~, \label{fixConstant}
\end{align}
where the subscript H denotes the HOMO of the KS system.
In Sec.\ \ref{SEC:DD} we will show that Eq.\ \eqref{fixConstant} fixes the constant such that $v_\xc$ is equal to the xc potential of an 
ensemble allowing for fractional charges and evaluated at $N_0^-=\lim_{p\to0^-} N_0 + p$ and $N_0$ is the number of particles of 
the system. The weighted orbital sum on the last line of Eq.\ \eqref{kli} is typical of the KLI approximation and contains 
the information about the discontinuity of the xc potential as a function of the number of particles.\cite{hg13} Indeed, in Sec.\ \ref{SEC:DD}
we will show that the discontinuity is given by
\begin{align}
  \Delta_\xc  = \frac{1}{2}\ind{3}{r} \psi_\xc(\vr) |\nabla\phi_\mathrm{L}(\vr)|^2 - \ind{3}{r} v_{\xc,2}(\vr) |\phi_{\mathrm{L}}(\vr)|^2 ~, \label{DDmGGA}
\end{align}
where L denotes the LUMO of the KS system.

\subsection{tLDA potential} \label{SEC:tLDA_potential}

Let us now apply Eqs.\ \eqref{vxc_1} and \eqref{OEP_mGGA} to the tLDA functional [cf.\ Eq.\ \eqref{tLDA}] proposed in Sec.\ \ref{SEC:tLDA}. 
The density dependence enters only via a prefactor multiplying the energy density per particle while the latter is purely $\tau$-dependent.
Therefore we have 
\begin{align}
  v^\tLDA_{\xc, 1}(\vr) = \tilde{\epsilon}^\unif_\xc[\tau(\vr)] ~, \label{vxc1_tLDA}
\end{align}
and 
\begin{align}
  v^\tLDA_{\xc, 2}(\vr) = \ind{3}{r'} \psi^\tLDA_\xc(\vr') \frac{\delta \tau (\vr')}{\delta n(\vr)} ~. \label{vxc2_tLDA}
\end{align}
An approximate expression for $v^\tLDA_{\xc, 2}$ can be obtained by ignoring variations of the fictitious density, $n_\fictitious$, with respect to variations of the true density, i.e.,
we can write
\begin{align}
  v^\tLDA_{\xc, 2}(\vr) & = n(\vr) \ind{3}{r'} \frac{\partial \epsilon^\unif_\xc}{\partial n_\fictitious }(\vr') \frac{\delta n_\fictitious (\vr')}{\delta n(\vr)} \nn
  & \approx n(\vr) \frac{\partial \epsilon^\unif_\xc}{\partial n_\fictitious}(\vr) ~. \label{vxc2_tLDA_approximate}
\end{align}
In this way the potential still carries a nonlocal density dependence via the kinetic-energy density, but may not be nonlocal enough to
account for features such as the derivative discontinuity.

Considering only exchange, we can write the tLDA potential explicitly, since we know the exact form [cf.\ Eq.\ \eqref{exxLDA}],
\begin{align}
  v^\tLDA_{\exch, 1}(\vr) = - \frac{3}{4 \pi} \left(10 \pi^2 \tau(\vr) \right)^{1/5} ~, \label{vx1_tLDA}
\end{align}
and
\begin{align}
  \psi^\tLDA_\exch(\vr) = - \frac{1}{2 \pi} 3 \pi^2 n(\vr) \left( 10 \pi^2 \tau(\vr) \right)^{-4/5} ~. \label{psix_tLDA}
\end{align}
Using the approximation above of setting $\delta n_\fictitious/\delta n = 1$ we find the approximate expression
\begin{align}
  v^\tLDA_{\exch, 2}(\vr) & \approx - \frac{1}{4 \pi} 3 \pi^2 n(\vr) \left(10 \pi^2 \tau(\vr) \right)^{-2/5} ~. \label{vx2_tLDA}
\end{align}

\section{Derivative discontinuity} \label{SEC:DD}

The discontinuous change in the derivative of the xc energy, when crossing integer particle numbers,
is an important property of the exact xc functional. It is directly responsible for correcting the 
too small KS gap, and indirectly for accurately breaking chemical bonds.\cite{pplb82} In the latter
case it was shown that the xc potential develops a step, or a constant shift across one of the atoms, in a 
stretched molecule. The step aligns the chemical potentials of the fragments and remains finite even at infinite 
separation. In order to capture such a step the xc potential must have a highly nonlocal 
dependence on the density.

In order to investigate the derivative discontinuity for $\tau$-dependent functionals we first have to generalize the theory to allow for 
densities that integrate to a non-integer number of particles. For an average particle number $N=N_0+p$ where $N_0$ is an integer
and $0<p<1$ this can be achieved by introducing an ensemble of the form
\begin{align}
  \hat{\gamma}^> = (1 - p) |\Psi_{N_0}\ket \bra \Psi_{N_0}| + p |\Psi_{N_0+1}\rangle\langle \Psi_{N_0+1}| ~, \label{PPLBensemble+}
\end{align}
where $\Psi_k$ is the ground-state wave function with $k$ particles. Similarly, for $N = (N_0 -1) + p$, we have
\begin{align}
  \hat{\gamma}^< = (1 - p) |\Psi_{N_0-1}\ket\bra \Psi_{N_0-1}| + p |\Psi_{N_0}\ket\bra \Psi_{N_0}| ~. \label{PPLBensemble-}
\end{align}

The exact ensemble ground-state energy $E(N)$ will consist of straight line segments between the values at the integers, and 
the slope on the $-/+$ side of $E(N_0)$ is equal to the negative of the ionization energy and electron affinity, respectively.
Most functionals in DFT--when extended to ensemble densities--do not reproduce this behavior which is due to the lack
of a derivative discontinuity.

Within KS DFT the ensemble density is calculated from the KS system. Since the {\em same} KS system should be 
used to calculate the density of both members in the ensemble we have to use the ensemble xc potential, 
i.e., a potential $v_\xc=\delta E_\xc[n]/\delta n$ where $n$ can be a density integrating to a non-integer number of 
particles. It is important to notice that this potential does not reproduce the densities of the individual ensemble members 
simultaneously, but it does reproduce the total ensemble density. 

For $N > N_0$ we can determine the KS ensemble density as 
\begin{align}
  n(\vr) & = (1 - p) n_{N_0}(\vr) + p n_{N_0 + 1}(\vr) \nn
  & = \sum_{i=1}^{N_0} |\phi_i(\vr)|^2 + p | \phi_\mathrm{L}(\vr)| ^2 ~, \label{n+} 
\end{align}
where we note that the KS orbitals also depend on $p$. In the same way the gradient of the ensemble 
density is given by
\begin{align}
  \nabla n(\vr) & = (1 - p) \nabla n_{N_0}(\vr) + p \nabla n_{N_0 + 1}(\vr) \nn
  & = \sum_{i=1}^{N_0} \nabla |\phi_i(\vr)|^2 + p \nabla |\phi_\mathrm{L}(\vr)|^2 ~. \label{nabla_n+}
\end{align}
For $\tau$-dependent functionals we also need the ensemble KS kinetic-energy density,
\footnote{Note that we are employing the KS system at integer particle number $N_0$ and $N_0+1$, respectively, in order to write the first line of Eq.\ \eqref{tau+}.}
\begin{align}
  \tau(\vr) & = (1 - p) \tau_{N_0}(\vr) + p \tau_{N_0 + 1}(\vr) \nn
  & = \frac{1}{2} \sum_{i=1}^{N_0} |\nabla \phi_i(\vr)|^2 + p \frac{1}{2} |\nabla \phi_\mathrm{L}(\vr)|^2 ~. \label{tau+}
\end{align}

The derivative with respect to the particle number $N$ is equal 
to the derivative with respect to $p$ and we define the so-called Fukui functions in the limit $N \to N_0^+$: 
\begin{subequations} \label{fukuiFunctions}
  \begin{align}
    f^+(\vr) \equiv \left. \frac{\partial n(\vr)}{\partial N} \right|_+ & = |\phi_\mathrm{L}(\vr)|^2 + \left.\sum_{i=1}^{N_0}\frac{\partial |\phi_i(\vr)|^2}{\partial N}\right|_+ ~, \label{f_n} \\
    \nabla f^+(\vr) \equiv \left. \frac{\partial \nabla n(\vr)}{\partial N} \right|_+ & = \nabla|\phi_\mathrm{L}(\vr)|^2\nn
    & {} + \left. \sum_{i=1}^{N_0}\frac{\partial \nabla|\phi_i(\vr)|^2}{\partial N} \right|_+ ~, \label{f_nabla_n}  \\
    f^+_\tau(\vr) \equiv \left. \frac{\partial \tau(\vr)}{\partial N} \right|_+ & = \frac{1}{2} |\nabla\phi_\mathrm{L}(\vr)|^2 \nn
    & {} + \left. \frac{1}{2} \sum_{i=1}^{N_0}\frac{\partial |\nabla\phi_i(\vr)|^2}{\partial N} \right|_+ ~. \label{f_tau} 
  \end{align}
\end{subequations}
Similar quantities can straightforwardly be defined for $N<N_0$ and $N\to N_0^-$ using the ensemble of Eq.\ \eqref{PPLBensemble-}. 

\begin{table*}[t]
  \caption{Ionization energy, $I = -\varepsilon_\mathrm{H}$, HOMO-LUMO gap, $\Delta_\ks$, and derivative discontinuity, $\Delta_\xc$, for various xc functionals.
    According to Eq.\ \eqref{gap} the electron affinity, $A$, is obtained by subtracting the sum of $\Delta_\ks$ and $\Delta_\xc$ from the ionization energy.
    Note that $A$ is expected to be zero (or small and negative) for closed shell atoms, expressing the fact that the anion is not stable. $\Delta_\xc$ is thus
    largely underestimated for the meta-GGAs studied in this work.}
  \begin{ruledtabular}
    \begin{tabular}{c|rrr|rrr|rrr|rrr|c}
      Atom & & LDA & & & tLDA & & & TPSS & & & VS98 & & Exp.\footnotemark[1] \\
      & $-\varepsilon_\mathrm{H}$ & $\Delta_\ks$ & $\Delta_\xc$ & $-\varepsilon_\mathrm{H}$ & $\Delta_\ks$ & $\Delta_\xc$ & 
      $-\varepsilon_\mathrm{H}$ & $\Delta_\ks$ & $\Delta_\xc$ & $-\varepsilon_\mathrm{H}$ & $\Delta_\ks$ & $\Delta_\xc$ & I \\
      \hline
      He & 0.5704 & 0.5711 & 0.0000 & 0.5460 & 0.5391 & -0.0071 & 0.5989 & 0.5995 & 0.0000 & 0.5939 & 0.5946 & 0.0000 & 0.9037 \\
      Be & 0.2057 & 0.1286 & 0.0000 & 0.1990 & 0.1236 & -0.0275 & 0.2111 & 0.1347 & 0.0074 & 0.2084 & 0.1408 & -0.0074 & 0.3426 \\
      Ne & 0.4980 & 0.4956 & 0.0000 & 0.4829 & 0.4556 & -0.0120 & 0.4963 & 0.4926 & 0.0003 & 0.5038 & 0.5033 & 0.0032 & 0.7945 \\
      Mg & 0.1754 & 0.1247 & 0.0000 & 0.1773 & 0.1264 & -0.0162 & 0.1748 & 0.1227 & 0.0036 & 0.1719 & 0.1201 & -0.0021 & 0.2808 \\
      Ar & 0.3823 & 0.3728 & 0.0000 & 0.3663 & 0.3318 & -0.0141 & 0.3829 & 0.3735 & 0.0009 & 0.3864 & 0.3789 & 0.0084 & 0.582
    \end{tabular}
  \end{ruledtabular}
  \footnotetext[1]{ From Refs.\ \onlinecite{atomicEnergies1, atomicEnergies2}. }
  \label{TAB:ionizationEnergiesXC}
\end{table*}

The difference between the derivatives of $E(N)$ with respect to $N$ around an integer is
equal to the difference of the ionization energy, $I$, and the electron affinity, $A$.
When the ensemble energy $E(N)$ is evaluated within KS DFT this difference is given by
\begin{align}
  I - A = \left.\frac{\partial E}{\partial N}\right |_{+} - \left.\frac{\partial E}{\partial N}\right |_{-} = \Delta_s+ \Delta_\xc ~. \label{gap}
\end{align}
$\Delta_s=\varepsilon_\mathrm{L} - \varepsilon_\mathrm{H}$ is the KS gap defined as the difference between the 
LUMO and the HOMO energy of the KS system. The true fundamental gap is thus equal to the KS gap plus a constant 
$\Delta_\xc$. This constant is given by, \cite{hg12,hg13}
\begin{align}
  \Delta_\xc = \left. \frac{\partial E_\xc}{\partial N} \right|_{+} - \int \!{\rm d}^3r \,v^-_{\rm xc}(\vr)f^+(\vr) ~. \label{deltaxc}
\end{align}
and known as the derivative discontinuity. $\Delta_\xc$ is also exactly equal to the discontinuous jump in the xc potential 
when passing an integer. It is important to keep track of the superscript $+/-$, because it signifies that the quantities are 
evaluated at $N\to N_0^{+/-}$, respectively. Equation \eqref{deltaxc} can easily be deduced by using the chain rule
\begin{align}
  \frac{\partial E_\xc}{\partial N} = \ind{3}{r} v_\xc(\vr)f(\vr) ~. \label{derexc}
\end{align}
First of all, evaluating this derivative at $N\to N_0^{-}$ yields
\begin{align}
  \left.\frac{\partial E_\xc}{\partial N}\right |_{-} = \ind{3}{r} v^-_\xc(\vr)f^-(\vr) ~. \label{xccond}
\end{align}
This equation does not allow for an arbitrary constant in $v_\xc$ and can therefore be used to fix the constant 
in the OEP equation at integer $N$. If the potential is fixed in this way, the HOMO of the KS system 
exactly corresponds to the negative of the ionization energy.\cite{Almbladh}
Secondly, if we evaluate the derivative at $N\to N_0^{+}$, we can write Eq.\ \eqref{derexc} as
\begin{align}
  \left.\frac{\partial E_\xc}{\partial N}\right |_{+} = \ind{3}{r} [ v^-_\xc(\vr)+\Delta_\xc ] f^+(\vr) ~, 
\end{align}
where we used that there is a constant jump by $\Delta_\xc$ when going from $v_\xc^-$ to $v_\xc^+$.
Using that $f^+$ integrates to one we arrive at Eq.\ \eqref{deltaxc}. 

In order to obtain Eq.\ \eqref{DDmGGA} of Sec.\ \ref{SEC:mGGA_OEP_KLI}, we now specialize 
the discussion to meta-GGAs, i.e., functionals of the form given by Eq.\ \eqref{excMGGA}. 
The derivative of $E_\xc$ with respect to $N$ in the limit $N\to N_0^+$ is given by
\begin{align}
  \left.\frac{\partial E_\xc}{\partial N}\right |_+ & = \ind{3}{r} \left\{ \epsilon_\xc(\vr) + n(\vr)\frac{\partial\epsilon_\xc}{\partial n}(\vr) \right\} f^+(\vr) \nn
  & {} + \ind{3}{r} n(\vr)\frac{\partial  \epsilon_\xc}{\partial \nabla n}(\vr)\nabla f^+(\vr) \nn
  & {} + \ind{3}{r} \psi_\xc(\vr) f^+_ \tau(\vr) ~. \label{dExcdN+}
\end{align}
From this expression we subtract
\begin{align}
  \ind{3}{r} v^-_\xc(\vr) f^+(\vr) & = \ind{3}{r} \left\{ \epsilon_\xc(\vr) + n(\vr)\frac{\partial\epsilon_\xc}{\partial n}(\vr) \right\}f^+(\vr) \nn
  & {} - \ind{3}{r} \left\{ \nabla \cdot n(\vr) \frac{\partial \epsilon_\xc}{\partial \nabla n}(\vr) \right\} f^+(\vr) \nn
  & {} + \ind{3}{r} v^-_{\xc,2}(\vr)f^+(\vr) \label{vxc-f+},
\end{align}
where we have used Eqs.\ \eqref{vxc_1} and \eqref{vxc_2} of Sec.\ \ref{SEC:mGGA_OEP}.
The terms in the first lines of Eqs.\ \eqref{dExcdN+} and  \eqref{vxc-f+} cancel trivially and we conclude that there is no discontinuity 
in functionals depending only on the density. The terms in the second lines of Eqs.\ \eqref{dExcdN+} and \eqref{vxc-f+}, due to the gradient, 
also cancel after a partial integration. However, this is under the assumption that $n\frac{\partial  \epsilon_\xc}{\partial \nabla n} f^+$ vanishes at the boundary. 
It was recently shown that with a diverging xc density one can construct a GGA exhibiting a derivative discontinuity,\cite{ak13} 
which is consistent with our analysis. Finally, the terms in the third lines of Eqs.\ \eqref{dExcdN+} and  \eqref{vxc-f+}, due to the $\tau$-dependence, will,
in general, not cancel and we find
\begin{align}
  \Delta_\xc = \ind{3}{r} \psi_\xc(\vr) f^+_\tau(\vr) - \ind{3}{r} v^-_{\xc,2}(\vr) f^+(\vr) ~. \label{DD_tau}
\end{align}
In order to calculate $\Delta_\xc$ in practice we need to first calculate $v^-_{\xc,2}$. 
This is done via the OEP equation \eqref{OEP_mGGA} in Sec.\ \ref{SEC:mGGA_OEP}.
This equation can, however, only determine the potential up to a constant. In order to fix the constant we have to 
impose Eq.\ \eqref{xccond}, which for $\tau$-functionals becomes
\begin{align}
  0 = \ind{3}{r} \psi_\xc(\vr) f^-_\tau(\vr) - \ind{3}{r} v^-_{\xc,2}(\vr) f^-(\vr) ~. \label{fixConstant_tau}
\end{align}
Due to the OEP equation the orbital relaxation terms of the Fukui functions cancel. 
Consequently, we can simply replace $f^- = |\phi_\mathrm{H}|^2$, $f^+ = |\phi_\mathrm{L}|^2$ and $f_\tau^- = 1/2|\nabla\phi_\mathrm{H}|^2$, $f_\tau^+ = 1/2|\nabla\phi_\mathrm{L}|^2$, respectively,
and we have proved Eqs.\ \eqref{fixConstant} and \eqref{DDmGGA}.

\section{Results: Spherical atoms} \label{SEC:results}

We have implemented the KS and OEP equations for spherical atoms using a
numerical approach based on cubic splines as radial orbital basis functions.\cite{splines} 
This approach has shown to be very efficient and accurate for the solution of OEP-type equations.\cite{hvb07,hvb09} 
The splines are defined over a cubic mesh with a cutoff radius of $r_\mathrm{max} = 80$, beyond which the wave functions
vanish. All results have been obtained using $\sim 200$ cubic splines ($\sim 100$ splines for $r < 10$)
and functionals, except for the tLDA and VSX, have been imported from Libxc.\cite{LIBXC}

\begin{table*}[t]
  \caption{ionization energy, $ I = -\varepsilon_\mathrm{H}$, HOMO-LUMO gap, $\Delta_\ks$, and derivative discontinuity, $\Delta_\mathrm{x}$, for various exchange functionals.}
  \begin{ruledtabular}
    \begin{tabular}{c|rrr|rrr|rrr|rrr}
      Atom & & tLDAx & & & TPSSx & & & VSX & & & EXX & \\
      & $-\varepsilon_\mathrm{H}$ & $\Delta_\ks$ & $\Delta_\mathrm{x}$ & $-\varepsilon_\mathrm{H}$ & $\Delta_\ks$ & $\Delta_\mathrm{x}$ & 
      $-\varepsilon_\mathrm{H}$ & $\Delta_\ks$ & $\Delta_\mathrm{x}$ & $-\varepsilon_\mathrm{H}$ & $\Delta_\ks$ & $\Delta_\mathrm{x}$ \\
      \hline
      He & 0.4932 & 0.4926 & -0.0022 & 0.5724 & 0.5730 & 0.0000 & 0.5892 & 0.5900 & -0.0001 & 0.9180 & 0.7597 & 0.2439 \\
      Be & 0.1638 & 0.1198 & -0.0231 & 0.1869 & 0.1300 & 0.0081 & 0.1968 & 0.1277 & -0.0013 & 0.3092 & 0.1312 & 0.2428 \\
      Ne & 0.4282 & 0.4172 & -0.0079 & 0.4609 & 0.4570 & 0.0003 & 0.4888 & 0.4749 & -0.0175 & 0.8507 & 0.6585 & 0.2993 \\
      Mg & 0.1432 & 0.1178 & -0.0131 & 0.1510 & 0.1162 & 0.0026 & 0.1573 & 0.1167 & -0.0100 & 0.2530 & 0.1168 & 0.1814 \\
      Ar & 0.3187 & 0.3024 & -0.0099 & 0.3473 & 0.3382 & 0.0006 & 0.3674 & 0.3452 & -0.0203 & 0.5908 & 0.4300 & 0.2398 \\
    \end{tabular}
  \end{ruledtabular}
  \label{TAB:ionizationEnergiesX}
\end{table*}

\begin{figure*}[t]
  \includegraphics[width=8.5cm, clip=true]{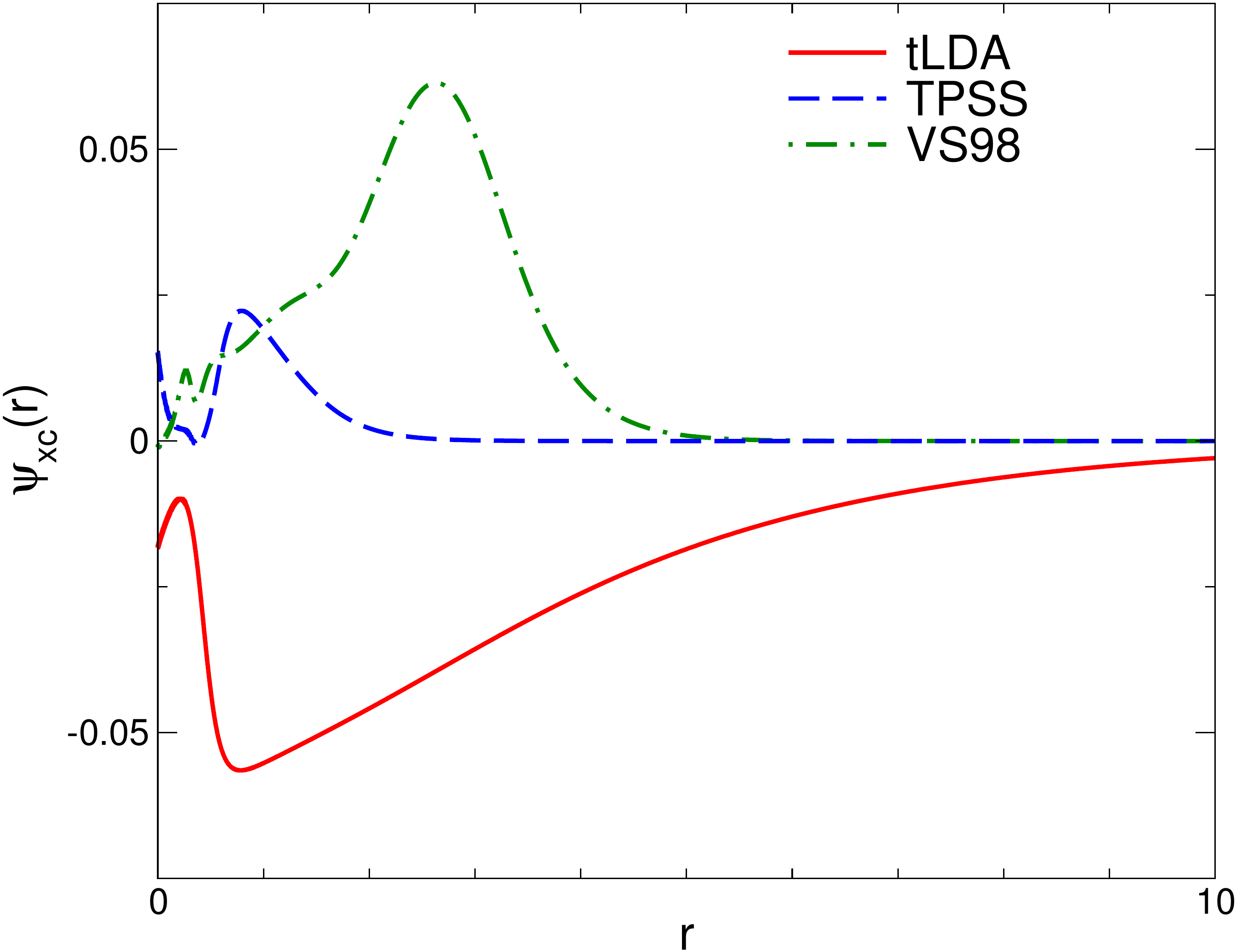}\hspace{5mm}
  \includegraphics[width=8.5cm, clip=true]{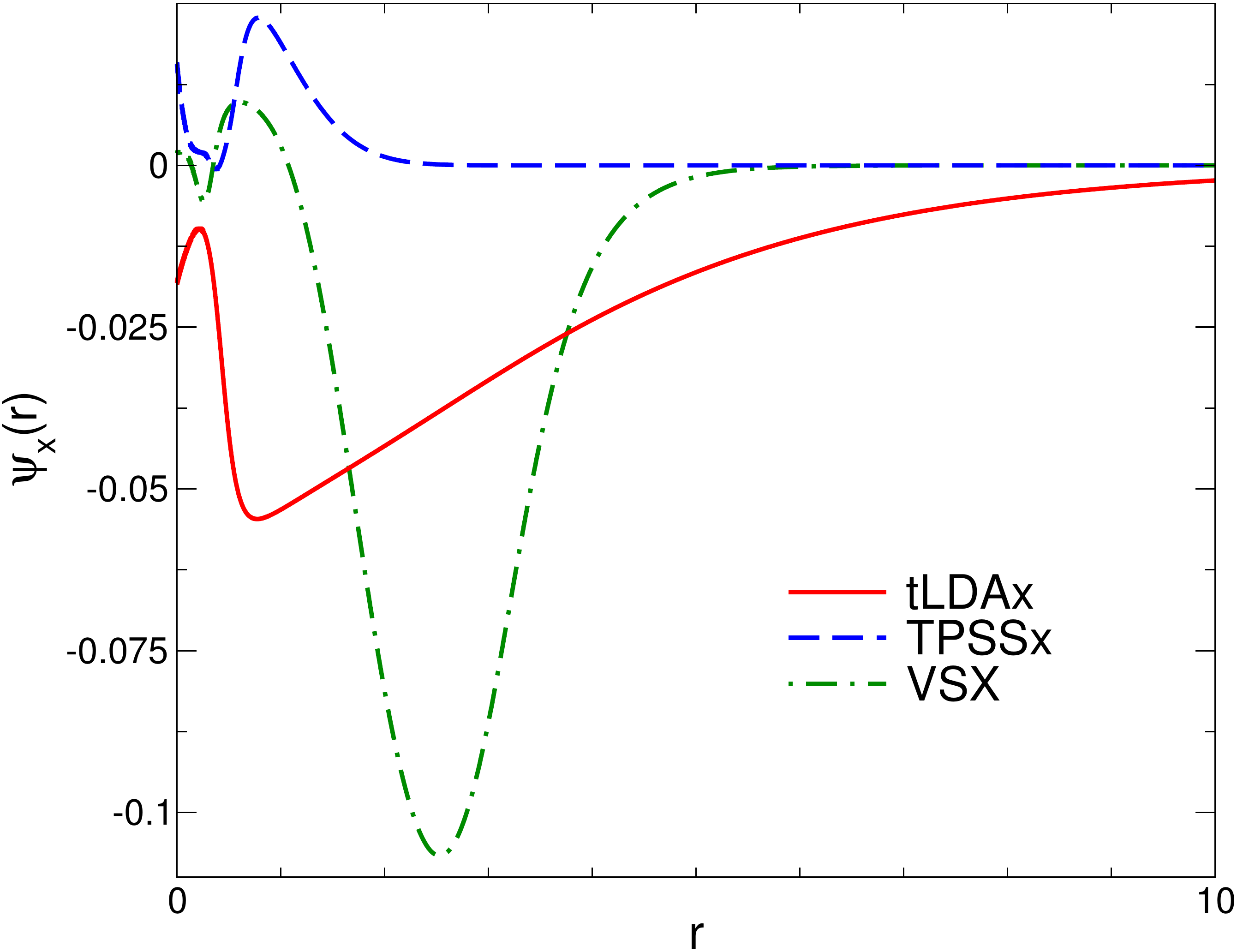}
  \caption{(Color online) Left panel: Plot comparing $\psi_\xc$ of Neon employing tLDA, TPSS, and VS98. \\
    Right panel: Plot comparing $\psi_\mathrm{x}$ of Neon using tLDAx, TPSSx, and VSX.}
  \label{FIG:NeonPsiPotentials}
\end{figure*}

The total energies for spherical atoms from Helium to Argon are shown in Table \ref{TAB:totalEnergyXC}, Sec.\ \ref{SEC:tLDA},
for the LDA, tLDA, the PBE~\cite{pbe96} GGA and two popular meta-GGAs functionals, the TPSS~\cite{tpss} and the VS98.\cite{vvs98}
For comparison we also show the EXX and the exact nonrelativistic total energies.
The simple modification of the LDA, proposed in Sec.\ \ref{SEC:tLDA}, improves the total energies compared to the original LDA, as has been already discussed in Sec.\ \ref{SEC:tLDA}. 
Furthermore, the tLDA supports one additional bound state for the noble gases Neon and Argon
and two additional bound states for the alkaline earth metals Beryllium and Magnesium. This has, however, to be compared to EXX, which yields a Rydberg series,
due to the $r^{-1}$ decay of the EXX potential.

In Table \ref{TAB:ionizationEnergiesXC} we show $I = -\varepsilon_\mathrm{H}$, the KS HOMO-LUMO gap, and the derivative discontinuity for the three studied meta-GGAs. 
The tLDA shifts the HOMO upwards and simultaneously the LUMO downwards. Accordingly, the KS gap is reduced compared to LDA.
In addition the derivative discontinuities in tLDA are negative for all atoms under investigation. This implies that the derivative discontinuity of tLDA shifts the
KS orbitals further down in energy. For TPSS the magnitude of the derivative discontinuity is much smaller, e.g., for Be the derivative discontinuity is $\sim 5$ percent
of the KS gap for TPSS and $\sim 20$ percent of the KS gap for tLDA. In contrast to tLDA all studied derivative discontinuities are positive in TPSS.
Interestingly we see similar trends in tLDA and TPSS regarding the magnitude of the derivative discontinuities, i.e.,
they decrease with the atomic number for the alkaline earth metals, while they increase with increasing
atomic number for the noble gases. The VS98 meta-GGA exhibits positive derivative discontinuities for the noble gases and negative derivative discontinuities for the alkaline earth metals, while 
the trend in absolute size of the derivative discontinuities is similar to tLDA and TPSS.

\begin{center}
  \begin{figure*}[t]
    \includegraphics[width=8.5cm, clip=true]{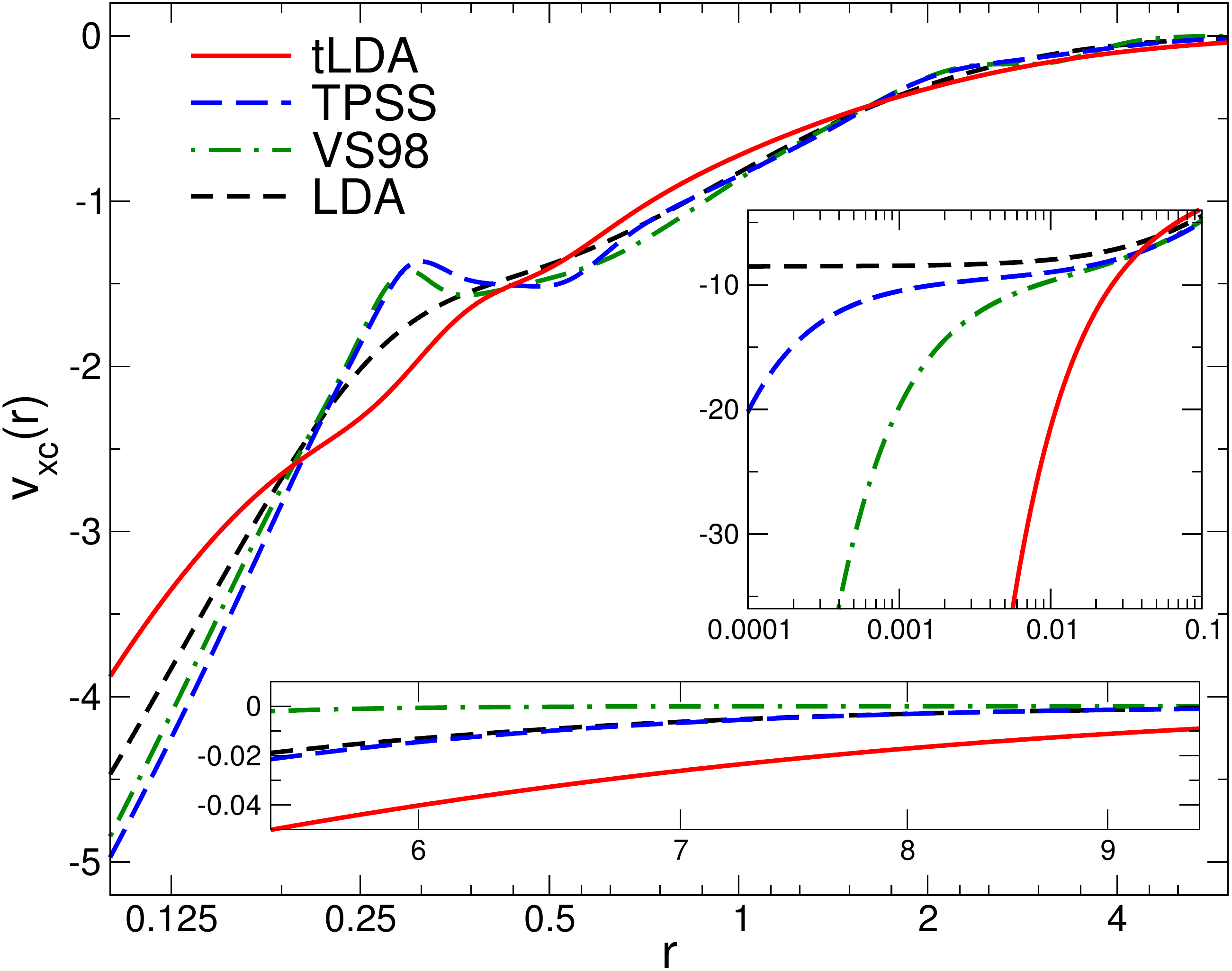}\hspace{5mm}
    \includegraphics[width=8.5cm, clip=true]{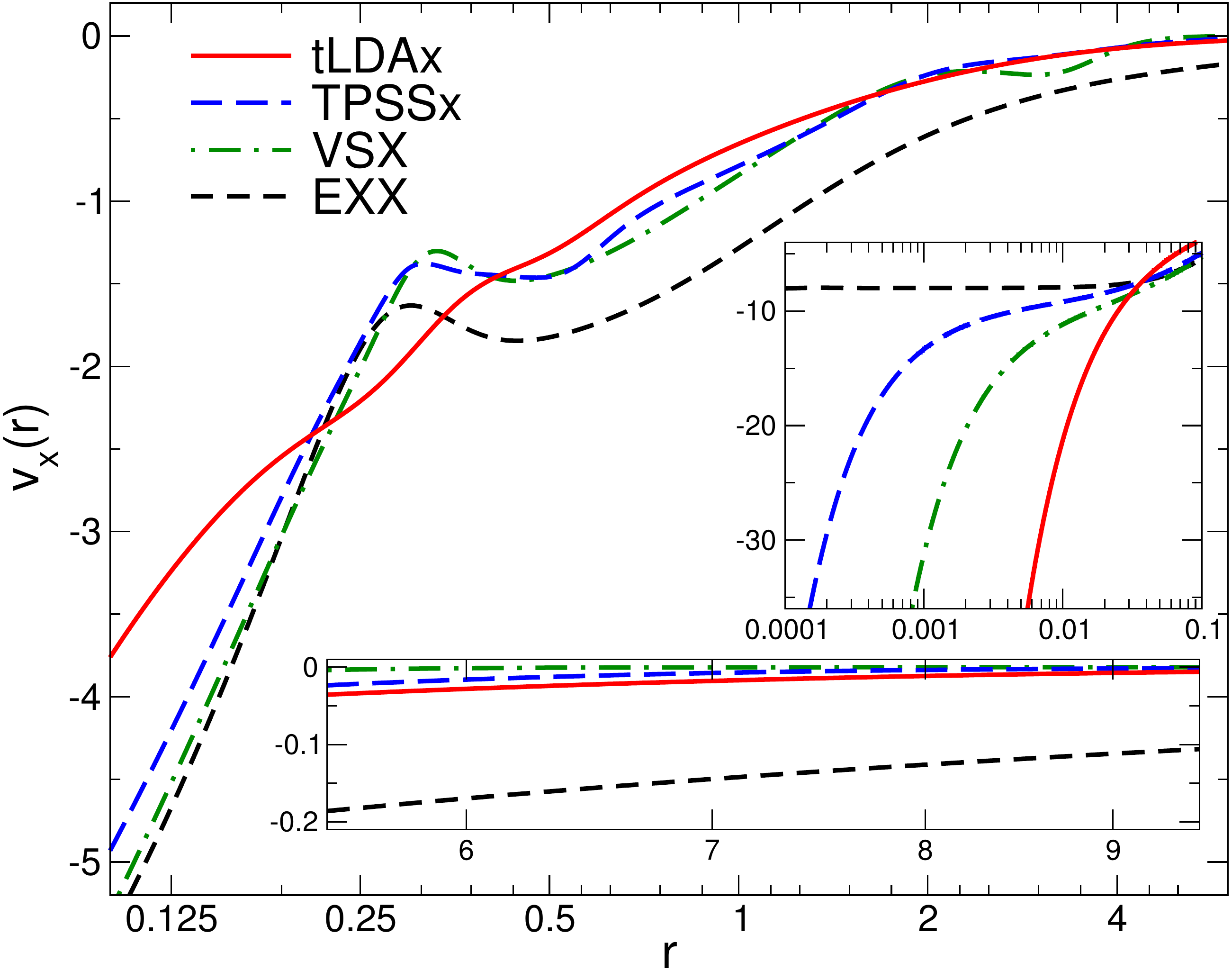}
    \caption{(Color online) Left panel: Plot comparing the selfconsistent xc potentials of Neon for tLDA, TPSS, VS98, and LDA. \\
      Right panel: Plot comparing the selfconsistent exchange potentials of Neon for tLDAx, TPSSx, VSX, and EXX.}
    \label{FIG:NeonPotentials}
  \end{figure*}
\end{center}

In order to investigate the derivative discontinuities for meta-GGAs further, we compare the ionization energy, $\Delta_\ks$,
the derivative discontinuity at the exchange level, where EXX provides the
``exact'' results. As can be seen in Table \ref{TAB:ionizationEnergiesX} both tLDAx and TPSSx show the same trends as in the xc results.
Moreover, we find that correlation is reducing the derivative discontinuity in tLDA, i.e., the derivative discontinuity becomes more negative.
The VSX approximation does not correspond to the exchange part of the VS98 functional, instead it is a
parametrization of the exchange form of VS98, which has been optimized to reproduce Hartree-Fock energies.\footnote{The corresponding parameters are given in Table VIII of Ref.\ \onlinecite{vvs98}.}
Similar to the tLDA all derivative discontinuities are negative for VSX. Note, however, that the magnitude of the derivative discontinuities for the
alkaline earth metals increases with the atomic number. For EXX the derivative discontinuity, $\Delta_\exch$, shifts all unoccupied KS levels above the zero of energy, i.e.,
it ``unbinds'' the KS Rydberg series mentioned earlier. 

In Fig.\ \ref{FIG:NeonPsiPotentials} we show $\psi_\xc$, i.e., the derivative of the xc energy density with respect to the kinetic-energy
density, for Neon. The left panel depicts $\psi_\xc$ for the tLDA, TPSS, and VS98. The right panel shows the
exchange $\psi_\exch$ for tLDAx (cf.\ Eq.\ \eqref{psix_tLDA} in Sec.\ \ref{SEC:tLDA_potential}), TPSSx, and VSX.
We observe that there is a correlation with the overall sign of $\psi_\xc$ and the sign of the corresponding derivative discontinuity, for both the exchange and the xc results.
In addition, the norm of $\psi_\xc$ seems to be a rough indicator for the size of $\Delta_\xc$. $\psi_\xc$ in TPSS is only appreciable close to the nucleus, while for
VS98/VSX it is biggest in the outer shell region ($r \sim 2$) of the atom. The global minimum of $\psi_\xc$ in the tLDA occurs approximately at the same position
as the global maximum of TPSS. Close to the nucleus the tLDA $\psi_\xc$ is roughly a mirror image of the TPSS $\psi_\xc$, however, it decays much slower than $\psi_\xc$ 
for TPSS and VS98/VSX .

Finally, we show in Fig.\ \ref{FIG:NeonPotentials} the xc potentials of Neon for xc (left panel) and the exchange (right panel).
They are obtained by solving the full OEP equation \eqref{OEP_mGGA}. The main plots show the potentials in the region $0.1 < r < 4.5$.
Both, TPSS and VS98/VSX, exhibit a pronounced shell structure around $r \sim 0.25$, which is similar to the shell structure of the
EXX potential (shown in the right panel) albeit shifted upwards. The tLDA potential also shows a shell structure, which is, however,
not as strong as the shell structure in the other two meta-GGAs. The right inset depicts the potentials in the vicinity of the nucleus ($10^{-4} < r < 0.1$).
While the LDA potential (shown in the left panel) and the EXX potential (shown in the right panel) have a finite value at the origin, the meta-GGA potentials
diverge. It is noteworthy that for TPSS the divergence due to the dependence on the gradient of the density and the divergence due to the
dependence on the kinetic-energy density are almost balanced. The bottom inset shows the asymptotic region of the potentials. We can clearly see
that the tLDA decays slower than the LDA, TPSS and VS98/VSX. This is due to the slow decay of $\psi_\xc$, which has been discussed previously.
The slower decay is responsible for binding additional states. In the bottom inset of the right panel we show the $r^{-1}$ decay of the EXX
potential for comparison.

\section{Conclusions} \label{SEC:conclusions}

In the presented work we have analyzed the derivative discontinuity and the
xc potential of meta-GGAs. We have proposed a simple meta-GGA, which is obtained
by replacing the density dependence in the LDA xc energy density per particle with a 
kinetic-energy density dependence via the Thomas-Fermi relation of the electron gas.
This so-called tLDA yields improved total energies of atoms through a largely improved exchange energy.

We have derived the OEP equation for meta-GGAs and provided an novel interpretation 
in terms of a linear response equation that turns a position-dependent mass 
into a local xc potential. By generalizing the theory to ensembles that allow for non-integer
number of particles, we have further derived an expression for the corresponding derivative
discontinuity.

We have explicitly evaluated the OEP potential and the derivative discontinuity for the 
tLDA. It supports an additional bound state but does not improve the ionization energy
and produces smaller KS gaps in comparison to the LDA. The functional exhibits a derivative discontinuity 
but it is two orders of magnitude smaller than its true value and has the wrong sign.
Using more sophisticated meta-GGAs, we have found that they exhibit a similar trend, i.e., 
the ionization energy and the KS gaps do not improve significantly. The discontinuity remains 
very small but is generally positive. 

Considering the importance of accurately capturing the discontinuity for describing charge-transfer,
bond dissociation and strongly correlated systems, this work gives valuable insights into how 
present meta-GGA functionals will perform in these situations. Our analysis
shows that meta-GGAs exhibit a derivative discontinuity, in contrast to local (LDA) and semi-local
(GGA) xc energy functionals. However, the derivative discontinuity is too small to correct
the KS gap towards the fundamental gap of the interacting system. Our result indicate that
a stronger dependency on the kinetic-energy density may be needed in order to obtain reasonable
derivative discontinuities from meta-GGAs.

\begin{acknowledgments}
F. G. E. gratefully acknowledges support from DOE under Grant No. DE-FG02-05ER46203.
We would like to acknowledge fruitful discussions with G. Vignale and E. K. U. Gross.
\end{acknowledgments}

\bibliography{tLDA}

\end{document}